\documentclass[12pt,preprint]{aastex}

\def \etal{{\em et al.}}

\newcommand\kms{km~s$^{-1}$}

\shorttitle{Ram Pressure Stripping in NGC 4476}
\shortauthors{Lucero \etal.}

\begin{document}


\title{Ram Pressure Stripping in the Low Luminosity Virgo Cluster Elliptical Galaxy NGC 4476}
    

\author{D. M. Lucero and L. M. Young}
\affil{Physics Department, New Mexico Institute of mining and Technology,
Socorro, NM 87801} 
\email{drundle@nmt.edu,lyoung@physics.nmt.edu}
\and
\author{J.H. van Gorkom}
\affil{Department of Astronomy, Columbia University
, 550 West 120th Street
New York, New York 10027 (USA)}
\email{jvangork@astro.columbia.edu}
\nocite{*}
\begin{abstract}
We present a deep VLA search for HI emission from the low-luminosity
Virgo Cluster elliptical galaxy NGC 4476, which contains
1.1$\times$10$^8$$M_{\odot}$ of molecular gas in an undisturbed disk
in regular rotation.  No HI was detected.  The rms noise in the final
image corresponds to a 3$\sigma$ column density sensitivity of
1.2$\times$10$^{20}$ cm$^{-2}$ at the position of NGC 4476, averaged
over the 4~kpc beam.  The total HI mass is less than
1.5$\times$10$^{7}$ $M_{\odot}$. If we compare our HI upper limit to
the H$_2$ content, we find that NGC 4476 is extremely deficient in HI
compared to other galaxies detected in these two species.  The 
H$_2$/HI mass ratio for NGC 4476 is~ $>$ 7, whereas typical H$_2$/HI ratios for
elliptical galaxies detected in both HI and H$_2$ are $<$~2.  Based on
this extreme HI deficiency and the intra-cluster medium (ICM) density
at the projected distance from M87 we argue that either NGC 4476 has
undergone ram-pressure stripping while traveling through the Virgo
cluster core or its average molecular gas density is larger and its
interstellar UV field is smaller than in typical spiral galaxies.
NGC 4476
is located 12\arcmin~ in projection from M87, which causes extreme
continuum confusion problems. We also discuss in detail the techniques
used for continuum subtraction.  The spectral dynamic range of our
final image is 50,000  to 1.   
\end{abstract} 

\section{Introduction}

In recent years it has become clear that the majority of elliptical
galaxies contain small amounts of cold interstellar gas and dust
(eg. Knapp \etal~1989; Colbert \etal~2001; Sadler \etal~2000).  Fifty
to seventy percent of ellipticals show HI emission in amounts
M(HI)/L$_B$$\geq$10$^{-3}$ (Huchtmeier, Sage, $\&$ Henkel 1995).  
The origin of the neutral atomic gas is still being debated; possibilities
include accretion of gas-rich companions, fallback of gas after the major
merger event that created the elliptical, secondary infall, and (especially
for the smaller systems) stellar mass loss (e.g.\ van Gorkom \&
Schiminovich 1997; Oosterloo et al.\ 1999).

Work done by Quillen \etal~(1992), Wiklind \etal~(1997), and Young
(2002) has shown that molecular gas in elliptical galaxies can be
found in flat disks in regular rotation about the central nucleus.  HI
studies in low-luminosity elliptical galaxies show that their HI
systems fundamentally resemble those of spiral galaxies (Sadler
\etal~2000).  The primary difference between the HI disks in
low-luminosity ellipticals and those in spiral galaxies is that the
surface density of the HI disk is a factor of 5 to 10 lower in
ellipticals (Oosterloo \etal~1999).  Typical $M_{H_2}$/M$_{HI}$ for
spiral and elliptical galaxies are 2 (Kenny $\&$ Young 1989, Thronson
\etal~1989) and 0.8 (Lees \etal~1990,  Wiklind \etal~1995), respectively.

It is well established that cluster spirals are deficient in HI
(Solanes \etal~2001).  HI imaging of a few clusters has shown that the
HI disks have smaller diameters in galaxies closer to the centers of
clusters (Warmels 1988, Cayatte \etal~1990; Bravo-Alfaro \etal~2000).
Studies of the Virgo Cluster show extreme HI deficiencies for galaxies
within 6$^{\circ}$ of the cluster center.  Inside this region, six
spiral and lenticular (S0) galaxies have $M_{H_2}$/M$_{HI}$ ratios of
10 and greater (see Table 1).  The fate of the HI in these galaxies
has been the subject of  much speculation.  Two main processes are
thought to affect the gas content of galaxies in clusters: (i) ICM-ISM
interactions (Gunn $\&$ Gott 1972; Nulsen 1982; Valluri $\&$ Jog 1990;
Schulz $\&$ Struck 2001; Vollmer \etal~2001), and 
(ii) gravitational interactions, including galaxy-galaxy encounters and 
cluster-galaxy encounters (e.g.\ Mihos 2004; Moore, Lake, \& Katz 1998; Henriksen \&
Byrd 1996).  ICM-ISM interactions would affect the gas in a galaxy but not
the existing stellar populations, whereas gravitational interactions would
affect both gas and stars.
Since the gas
disks in low-luminosity ellipticals resemble those of spirals, it is
very likely that the cluster environment will have the same effects on
HI disks in both spirals and ellipticals.  In this paper we investigate this type of alteration of the ISM in the low-luminosity elliptical NGC 4476.

Spirals which are deficient in HI but not in H$_2$ may have enhanced
H$_2$/HI mass ratios.  The largest H$_2$/HI mass ratios in Virgo
spirals are listed in Table \ref{tbl-1}, and so far no elliptical
galaxies are known to have mass ratios as large as these.  Of the
ellipticals which are detected in CO, Wiklind \etal~(1995), Lees
\etal~(1991), and Georgakakis \etal~(2001) find only five with
$M_{H_2}$/M$_{HI}$ of at least 3, and these ratios are all based on HI
upper limits except for NGC 2623 (see Table 2).  The remaining six
entries in Table 2 are H$_2$ non-detections so the H$_2$/HI ratios
calculated for these galaxies are upper limits.  The lenticulars which
are studied in detail by Welch and Sage (2003) have H$_2$/HI in the range 0.01 to 1.5 which may be related to the fact that their sample excludes Virgo cluster members.   

In this paper we present a search for HI emission from the
low-luminosity Virgo elliptical NGC 4476.  The main challenge in this
observation was to achieve good spectral dynamic range, since M87 is a
strong radio source ($\sim$220 Jy at 1.4 GHz).  In Sec. 3 we describe
the method used to achieve optimal bandpass calibration and continuum
subtraction.  No HI was detected in the entire 40\arcmin~ by
40\arcmin~ by 647 km s$^{-1}$ data cube.   In Sec. 4 we discuss the
implications of this null result, and provide evidence that the HI gas
has been stripped by the ICM surrounding M87 and/or the physical
conditions of the ISM in NGC 4476 are very different than observed for
normal stripped spirals. In Sec. 5 we discuss the effects of ram
pressure on the star formation in NGC 4476.    In Sec. 6 we present our conclusions.

\section{NGC 4476}\label{4476}

NGC 4476 is a small elliptical galaxy 12\arcmin~from M87.
Measurements of the surface brightness fluctuations yield a distance
of 17.2$\pm$1.4 Mpc (Tonry \etal~2001) and an absolute magnitude of
$-$18.17 assuming a distance modulus of 31.18 $\pm$ 0.17.  The stellar
kinematics are normal for a low-luminosity elliptical galaxy, with a
central velocity dispersion of 132~km~s$^{-1}$ (Simien $\&$ Prugniel
1997).  Photometric observations show that NGC 4476 follows an
r$^{1/4}$ law throughout (Prugniel, Nieto, and Simien 1987).  Dust is
clearly seen on optical images, showing the presence of a nuclear dust
ring (Tomita \etal~2000). 

VLA 20 cm continuum observations of NGC 4476 in the B configuration
yield an upper limit of 5$\times$10$^{-30}$ W m$^{-2}$ Hz$^{-1}$ or
0.5 mJy (Owen 2001, private communication).  When compared to the IRAS
observations (Knapp \etal~1989), the radio continuum observations
indicate that NGC 4476 is an order of magnitude underluminous in the
radio (q $>$ 3.3; Lucero \& Young in prep.) compared to the radio-FIR correlation (Helou \etal~1985).  

The galaxy appears to be optically undisturbed, and recent
observations with the BIMA mm interferometer show a flat molecular
disk in regular rotation which also appears to be undisturbed (Young
2002).  H$_2$ masses are calculated using galaxy distances and a
``standard'' CO/H$_2$ conversion factor of 3.0$\times$10$^{20}$
cm$^{-2}$(K~km~s$^{-1}$)$^{-1}$.  With this conversion factor, H$_2$
masses are related to CO fluxes S$_{CO}$  by:
\begin{equation}
M(H_2)=(1.18\times10^4\mathrm{M_\Sun})D^2S_{CO}
\end{equation}
Where D is the distance in Mpc and S$_{CO}$ is the CO flux in
Jy~km~s$^{-1}$. Using the CO flux from Young (2002), we calculate an
H$_2$ mass of 1.1$\times$10$^8$ $\mathrm{M_\Sun}$.  A previous search
for HI in NGC 4476 puts an upper limit of 0.8 Jy km s$^{-1}$
(approximately 6.1$\times$10$^7$ $\mathrm{M_\Sun}$) on the HI content
(Giovanardi, Krumm, and Salpeter 1983) or M$_{H_2}$/M$_{HI}$ $>$~2 and
M$_{HI}$/L$_B<$~0.02.  The present paper improves on that limit by a
factor of 5, putting NGC 4476 in the same class as HI deficient
spirals and ellipticals with typical ratios of M$_{H_{2}}$/M$_{HI}$
$>$ 10 (Knapp, Helou, $\&$ Stark 1987) and $>3$ (Georgakakis
\etal~2000; Lees \etal~1991; Wiklind, Combes, $\&$ Henkel 1995),
respectively. Some physical parameters of NGC 4476 can be found in
Table  3.

\section{HI Observations and Strategy}

The observations were made on 2002 January 5 with the VLA\footnote{The
  VLA is operated by the National Radio Astronomy Observatory, which
  is a facility of the National Science Foundation (NSF), operated
  under cooperative agreement by Associated Universities.} in it's D
configuration.  For a description of the VLA see Napier
\etal~(1983). We used a total bandwidth of 3.125 MHz to cover the
  velocity range from 2284 km~s$^{-1}$ to 1637 km~s$^{-1}$ at a
  resolution of 10.4 km~s$^{-1}$.  The CO line is 200 km~s$^{-1}$ wide centered on 1960
km~s$^{-1}$ (Young 2002).  

Since we are dealing with a very large and complicated continuum
source ($\sim$200 Jy) in the close vicinity of NGC 4476, we employ a
strategy similar to that used by van Gorkom \etal~(1993) to obtain a
high dynamic range HI cube of 3C273.  During our ``on-source''
observations, the phase center of the array was placed on M87.  3C286
was chosen as the bandpass calibrator and absolute flux standard.
Fifteen minute scans were made of this source at the beginning and end
of our observations.  The flux density scale is based on an assumed
value of 14.885 Jy for 3C286 at 1414.19 MHz (Perley~1992).  The point
source J1254+116 was used as the phase calibrator, and six one and a
half minute scans were made every fifty minutes on this source.  The
total time on source was  4 hours.  The theoretical noise level of
these observations is ~0.5 mJy~beam$^{-1}$ in a 10.4~km~s$^{-1}$
channel in the center of the field or 0.71 mJy~beam$^{-1}$ at the
position of NGC 4476. The phase and gain are calibrated using the
average of the inner 75$\%$ of the bandpass (known as the channel zero
data).  Standard calibration procedures were followed to produce a
preliminary continuum image of the channel zero data.  Our ultimate
goal here is to remove all of the continuum associated with M87 and
expose all of the underlying HI.  Therefore, we need very high quality
 continuum calibration, which was achieved by self-calibrating the
channel zero data.  Self-calibration uses a clean component model of
point sources in the image to estimate corrections in the complex
antenna gains.  The final continuum image has a peak intensity of
80~Jy~beam$^{-1}$, and an rms noise of $\sim$~1~mJy~beam$^{-1}$, which
give a dynamic range (peak intensity divided by noise level) of 80,000 to 1.  The continuum image is shown in  Figure 1.

\subsection{Data Reduction}
\subsubsection{Bandpass Calibration}

Since M87 is a very strong continuum source it is necessary to do very careful bandpass calibrations.  
Some of the VLA antennas show a variation of the bandpass with time
due to a 3 MHz standing wave in the wave guide system.  
The standing wave pattern causes a time variable gain error of about 0.5$\%$ (Carilli 1991). It is necessary to calibrate out the time dependence of the gain when strong sources are present. 
After close inspection of the antenna-based complex bandpass functions, we found that our observations of 3C286 did not adequately model the time variability during the four hour interval between scans. 
 Also, extra noise was introduced due to the fact that 3C286 gives significantly lower signal to noise than M87.  

Therefore, we decided to use M87 as the bandpass calibrator.  We
found the signal to noise on M87 to be adequate to make
bandpass solutions every three minutes.  
A potential problem with this technique is that HI absorption against 
M87 or HI
emission close to the center of the field could cause spurious
features in the frequency-dependent gain solutions (van Gorkom
\etal~1993).  The spurious features would have velocities and linewidths
comparable to those of the spectral lines which caused them.
However, there are stringent upper limits on HI
absorption against M87 (N$_{HI}$ $<$ 5.0$\times$10$^{19}$ cm$^{-2}$,
assuming $\tau$ $<$ 2$\times$10$^{-4}$, and T$_s$=100 K; Dwarakanath
\etal~1994) and this effect should  be negligible. 

After the bandpass
calibration, some of the baselines still show a slight curvature of
$\sim$ 0.1$\%$ in amplitude across the 3 MHz bandwidth. 
The origin of this residual large scale curvature is not clear; the
features are too broad to be attributable to HI emission or absorption as
described above.

\subsubsection{Continuum Subtraction}

Since errors in the continuum subtraction dominate the thermal noise
in this dataset, we tried the following methods to  subtract the  continuum: 
\begin{enumerate}
\item The continuum was subtracted in the visibility  plane using the
  AIPS routine UVLIN.  
UVLIN operates on each visibility independently, fits a linear
spectral ``baseline'' to a selected group of channels, and then
  subtracts that fit from the  data.  
UVLIN is expected to leave residuals which are strongest at the position
of the strong continuum sources (Cornwell,   Uson, \& Haddad 1992). 
We tried this method using the channels 7-20 and 45-57, which  lie
outside the CO velocity range (there should be no HI line  emission here).
 Because of residual curvature in the bandpass-calibrated visibility
 spectra (see above) and because only the end channels were used, the
 rms noise in this image cube is relatively high and is dominated by
 residual  continuum features.
Mapping these data using natural weighting resulted
in a rms noise of 2.6 mJy~beam$^{-1}$. 
All of the noise levels  quoted in this section are
calculated in a 4\arcmin$\times$4\arcmin~ box at the position of NGC
4476, before the application of a primary beam correction.  

\item Again beginning from the original self-calibrated dataset, we
  tried running UVLIN to fit the entire usable frequency range (channels 10-55; 
1710-2178 km s$^{-1}$).  This
  method produced an rms of 1.4 mJy~beam$^{-1}$ at the position of NGC
  4476.
\item The continuum emission is very strong and complex on the shorter baselines,
  and much weaker on the longer  baselines.  
Therefore, an additional improvement in the noise levels is achieved
by restricting the map to the longer baselines (between 1 and 5
kilo-wavelengths).
The combination of UVLIN (fitting the entire usable frequency range) and
eliminating shorter baselines produces a data cube with an
rms noise  of  0.9 mJy~beam$^{-1}$.

\item Since UVLIN leaves residuals at the position of strong continuum
sources, we also tried the task UVSUB followed by UVLIN.  UVSUB
  subtracts from the visibility data a Fourier sum of the clean 
components estimated by deconvolution of the continuum image.
The use of UVSUB produced no improvement over steps 2 and 3 above.
\item
We believe that the difficulties in continuum subtraction are
primarily due to the residual 0.1\% curvature in the
bandpass-calibrated visibility spectra.  We thus attempted to remove the 
curvature
using the AIPS task IMLIN, which fits the spectrum of each spatial pixel in an image
cube (Cornwell, Uson, \& Haddad 1992).  We began with a cube made from a
continuum-subtracted dataset (step 1 above) and used IMLIN to fit a second
order polynomial to channels 7-20 and 45-57.
This procedure brought the noise level down to  1.0 mJy~beam$^{-1}$.  

\item We also tried IMLIN alone, fitting a second order polynomial over the entire usable channel range
  (10-55).  This method produced a rms of $\sim$ 1.4
  mJy~beam$^{-1}$, comparable to that produced by using a linear fit
  with UVLIN over the entire channel range.  Leaving out the
  baselines shorter than 1 kilowavelength produced a noise
  level of  0.9 mJy~beam$^{-1}$.  
\end{enumerate}
Steps 3, 5, and 6 produced non-primary beam corrected HI cubes of
comparable quality.  All of these HI cubes were imaged neglecting
baselines $<$~1 k$\lambda$ and have a beam size of 46.7\arcsec $\times$
42.1\arcsec\ (3.9 $\times$ 3.5 kpc).
By leaving out spacings less than 1 k$\lambda$ we have reduced
sensitivity to structures larger than 3.4\arcmin.  However, the CO
emission is only 30\arcsec~ in diameter,
so the lack of short spacings will only present a problem if the 
HI is distributed on size scales 7 times larger than the CO.
HI upper limits in the following sections are derived from the cube(s) with 
noise level 1.0 mJy~beam$^{-1}$, which is still twice the theoretical noise.  
The primary beam correction caused the thermal
noise to increase to 1.6 mJy~beam$^{-1}$  at the position of NGC  4476.

\subsection{HI Limits}

The primary beam-corrected HI spectrum 
is depicted in Figure 2.  Spatial smoothing has not been employed
since the CO emission is smaller than the 45\arcsec~ beam of the D
configuration.  An HI mass limit can be obtained by
integrating over the 200 km/s linewidth of the observed CO emission (20
channels, 2060-1860 km s$^{-1}$), and we find 0.015
Jy~km~s$^{-1}$.  The uncertainty in the sum of 20 independent channels
each of rms 1.6 mJy beam$^{-1}$ is 0.072 Jy~km~s$^{-1}$.  Thus, a
3$\sigma$ upper limit 
yields 0.21 Jy~km~s$^{-1}$ for the HI content of NGC 4476.
The corresponding mass limit is 
$< 1.5\times 10^7$ M$_\odot$ for an unresolved source, and the 
column density limit is $< 1.2 \times 10^{20}$ cm$^{-2}$.

Alternatively, if we choose not to assume that the HI covers the same
velocity range as the CO,
we may consider a six sigma limit of
the flux from only one channel.  Recall that the primary beam
corrected rms noise in one 10 km~s$^{-1}$ channel is 1.6 mJy
beam$^{-1}$. If we use this approach we get an individual channel
limit of 0.096 Jy~beam$^{-1}$~km~s$^{-1}$, which gives an HI column
density limit of $<$~5.4$\times$10$^{19}$~cm$^{-2}$ or a mass limit of
$<$~6.7$\times$10$^{6}$$\mathrm{M_\Sun}$. These upper limits put a
lower limit to the H$_2$/HI mass ratio of about 10 (ranging from 7 to 16
for the two different methods). H$_2$/HI mass ratios of this magnitude make
NGC 4476 the most HI deficient elliptical galaxy which is also known
to have molecular gas.  The absence of HI emission in NGC 4476 is
surprising, given that a photo-dissociated envelope should form around
the molecular gas (Blitz ~1993; Savage \etal~1977; Lada  \etal~1988).

\section{Discussion}

There are several possible reasons for the observed HI deficiency in NGC
4476.  The HI may have been removed through a tidal interaction with
another galaxy or by the ICM in the form of ram pressure
stripping.  As NGC~4476 is close to M87, there are many galaxies in the
vicinity which are candidates for gravitational interactions with NGC 4476.
The two closest in projection and in luminosity are NGC 4478 (4\arcmin\
away, $\Delta v = 610$ \kms) and NGC 4486B (11\arcmin, $\Delta v = 415$
\kms).  Both of these galaxies show evidence of tidal truncation and NGC
4478 also shows an isophote twist (van den Bosch \etal\ 1994).  However, it
is usually assumed that their tidal features are due to interactions with
M87 rather than with NGC~4476.  NGC~4476 itself shows no hint of disruption
either in the optical or in CO emission (van den Bosch \etal\ 1994; Young
2002), but the detectability of tidal features does depend somewhat on the
geometry of the interaction.  We can neither confirm nor disprove the
hypothesis that the extreme HI deficiency in NGC 4476 is due to a
gravitational encounter with another galaxy.  However,
consideration of ram pressures shows that it
is highly likely that NGC 4476 has suffered some ram pressure stripping.

\subsection{Ram pressure stripping}

The Virgo Cluster can be broken up into two major sub-clusters.  These
sub-clusters are centered on the two brightest galaxies in the
cluster, M87 and NGC 4472 (Ferrarese \etal~2000).  The distance to the
M87 sub-cluster has been determined to be 18 Mpc by the use of Cepheid
variables (Ferrarese \etal~2000).  The distances to M87 and NGC 4476
have also been determined to be 16.1$\pm$1.2 Mpc and 17$\pm$1.4 Mpc,
respectively, using the surface brightness fluctuation (SBF) method
(Tonry \etal~2001).  Therefore, we start our analysis by assuming that
NGC 4476 is a Virgo Cluster member, and we adopt a distance of 17 Mpc
for both M87 and NGC 4476.  M87 and NGC 4476 are 12\arcmin~apart which
translates to a separation of  60.0~kpc. 

The limits by Tonry (2001) locate NGC 4476 well within the Virgo
Cluster.  Its high radial velocity, 1978 km~s$^{-1}$, makes it likely
that it is in the central region of Virgo since a galaxy on a radial orbit
through the cluster reaches its highest velocity in the central region
(Solanes \etal~2001, Giovanelli $\&$ Haynes 1985, Dressler 1986,
Vollmer \etal~2001).  Given that, and given the very small projected
distance to M87, we can now make some estimates as to how close together
the galaxies really are and how effective ram pressure stripping will be 
at that separation.  

The gas in the galactic disk of NGC 4476 will be removed if the ram
pressure of the ICM is greater than the restoring gravitational force
per unit area (Gunn and Gott 1972).  This relationship is  given by:
\begin{equation}
{\Sigma_{gas}}{V_{rot}^{2}}R^{-1}=\rho_{ICM}V^{2}_{gal} 
\end{equation}
 where $\Sigma_{gas}$, V$_{rot}$, R, $\rho_{ICM}$, and V$_{gal}$ are
 the mass density of the gas disk, the rotational velocity of the gas,
 the radius of the gas disk, the ICM density, and the velocity of the
 galaxy through the ICM, respectively.  $\Sigma_{gas}$ is
 defined as:
\begin{equation}
{\Sigma_{gas}}=M_{gas}/A_{disk}
\end{equation}
where M$_{gas}$ and  A$_{disk}$ are the mass of the gas in the disk
and the surface area of the gas disk, respectively.  ROSAT observations
(B\"ohringer \etal~1994) show that the Virgo Cluster core is
filled with hot gas and that the M87 sub-cluster has the strongest X
ray emission.  According to Ferrarese (2000), to first order, M87 is
at the center of the X-ray emitting corona.  The ICM density at the 60
kpc distance from M87 is $n_e$ $\sim$ 3.60$\times$10$^{-3}$ cm$^{-3}$
(Nulsen $\&$ B\"ohringer 1995).  Assuming that the gas at this
distance is fully ionized and that the bulk of the mass comes from
the protons, the mass density is given  by:
\begin{equation}
\rho_{ICM}=1.15 n_e M_p
\end{equation}
where n$_e$ is the electron density, M$_p$ is the proton mass, and
1.15 is the mean mass per electron (including a contribution from helium).  The
heliocentric velocities of NGC 4476 and M87 are 1978 km~s$^{-1}$
(Simien $\&$ Prugniel 2001) and 1307 km s$^{-1}$ (Smith \etal~2000),
respectively.  The velocity of M87 is $<$ 200 km~s$^{-1}$ off from the
mean velocity of the M87 subcluster (Ferrarese \etal~2000).
Therefore, if we subtract the radial velocities of NGC 4476 and M87 we
obtain a lower limit to the relative velocity of NGC 4476 with respect
to the ICM.  This gives a ram pressure of
3.1$\times$10$^{-12}$~N~m$^{-2}$ at 60 kpc from  M87.

The CO disk rotates at $\sim$ 105 km~s$^{-1}$, and has a radius,
R$_{CO}$, of 1.2 kpc.  The H$_2$ mass is $\sim$ 1.1$\times$10$^8$
$M_\odot$ (Young 2002), and therefore the gas surface density is 24
M$_\odot$~pc$^{-2}$.  This value is a lower limit to the total gas surface
density because it does not include HI or helium, and it
is also an average over the whole disk, whereas the H$_2$ may be clumpy as
in Cen A (Quillen \etal~1992).  The total restoring pressure exerted
on the molecular disk by gravity is 1.5$\times$10$^{-11}$~N~m$^{-2}$.
Therefore, the ram pressure exerted on NGC 4476 is not sufficient to
strip the observed CO disk at the galaxy's current projected distance
from M87.  In fact, for a relative galaxy-ICM velocity of 671 km~s$^{-1}$, the
galaxy must get within 10 to 20~kpc of the center of M87 before the
ICM densities are large enough, $\sim$ 3.5$\times$10$^{-26}$~g~cm$^{-3}$, 
to strip the observed molecular gas disk.  Alternatively, at 
the current projected distance from M87, the velocity of NGC 4476 through
the intracluster medium would need to be on the order of 1470 km~s$^{-1}$
for molecular gas to be stripped.

It is important to keep in mind that there are several uncertainties
in these calculations.   The true relative velocity of the galaxy and the
ICM is probably larger than 671 km~s$^{-1}$, which of course is only the
radial component.  The true separation of the galaxies is probably 
larger than the projected separation.  Finally, the inclination angle
between the disk's spin axis and its velocity through the ICM is unknown.
Kenney, van Gorkom, \& Vollmer (2004) have argued that stripping rates
should be proportional to $\cos^2 i$, so that stripping is much less
effective for galaxies proceeding edge-on through the ICM.  Schulz \&
Struck (2001) suggested that the dependence on inclination would not
be as strong as $\cos^2 i$ on long timescales due to viscous effects.
In short, the conditions which we estimate to be necessary for stripping
are only accurate to factors of a few.

Studies of spiral galaxies in clusters have revealed that the CO content in these galaxies does not
appear to depend on environment (Stark \etal~1986; Kenney and Young
1989).  Our results indicate that low-luminosity ellipticals probably
behave like spirals in this respect, i.e. it's difficult to strip the
molecular gas.  We know of one counterexample, however: 
a recent study of the ISM in the spiral
galaxy NGC 4522 indicates that some of the CO in the outer parts of
this galaxy has also been stripped (Kenney, Van Gorkom, $\&$ Vollmer  2004).

The radial distribution of CO in many Sc galaxies resembles an
exponential with a scale length similar to that of the starlight
(Young $\&$ Scoville 1991).  We also model the original
(pre-stripping) gas distribution of NGC 4476 as an  exponential: 
\begin{equation}
\Sigma_{gas}=\Sigma_0~\exp(-R/1.2~kpc)
\end{equation}
where $\Sigma_0$ is the present molecular surface density (24
M$_{\odot}$ pc$^{-2}$) and the scale length is equal to the effective
radius of the galaxy (15\arcsec $=$ 1.2 kpc, Simien $\&$ Prugniel
1997).  The maximum extent of the present CO disk is also 1.2 kpc.  In
this model, the present ram pressure exerted on NGC 4476 is sufficient to
strip all of its ISM at radii beyond  1.7 kpc.  
A velocity of 890 km~s$^{-1}$ would be sufficient to strip all ISM at radii
beyond 1.2 kpc, where the model disk's gas density drops to 9 M$_\odot$
pc$^{-2}$.

Our HI column density limits for NGC 4476 are certainly low enough that
stripping of the outer HI disk is plausible. Based on
position-velocity diagrams for other ellipticals, we assume that the
HI disk would have a flat rotation curve beyond the edge of the
molecular gas at $>$ 1.2~kpc (Young 2002; Osterloo, Morganti, $\&$
Sadler 1999). The HI column density limit of 1.2$\times10^{20}$
cm$^{-2}$ and assumed HI disk velocity (105 km s$^{-1}$) give a
restoring pressure of $<$ 6.0$\times$10$^{-13}$ N~m$^{-2}$ at a radius
of 1.2 kpc.  The ICM
density needed to equalize the ram pressure and restoring pressure
is only $n_e \approx 7\times 10^{-4}$ cm$^{-3}$.  
According to
Nulsen $\&$ B\"ohringer (1995), this ICM density is located a
distance $\approx$ 200~kpc from M87.  Therefore, we can say that
unless NGC 4476 has always remained farther than 200~kpc from
M87 (ICM densities at this distance are too low to remove any HI), the
stripping explanation is at least consistent with the observed low HI
column densities.  All of the ISM and ICM gas properties derived in
this section can be found in Table  4.

The distance estimates discussed in the early part of this section are
not accurate enough to specify whether NGC~4476 is in the foreground of
M87 and approaching the cluster core, or is in the background of M87 and
receding from it.  Using a
N-body/sticky-particle code to simulate the time variable ram pressure
that a galaxy would feel as it moves through the Virgo Cluster core,
Vollmer \etal~(2001) find that the maximum damage to a galaxy's HI
disk only becomes apparent long after closest approach to the 
cluster center.  Therefore, if the HI disk in NGC 4476 has indeed been
completely stripped, then it is very likely that the galaxy has
already passed through the cluster  center. 

\subsection{A central HI hole?}

The nondetection of HI emission from NGC~4476 was a bit of a surprise,
given the abundance of molecular gas in the galaxy.
In most of the HI-deficient Virgo spirals, the
inner disks remain intact and are detected both in HI and in CO
(Kenney $\&$ Young 1989).  The peak HI and H$_2$ column densities in these
galaxies are $\sim$ 5$\times$10$^{20}$ cm$^{-2}$ and $\sim$
$2\times 10^{21}$ cm$^{-2}$, respectively (Cayatte \etal~1994;
Kenney $\&$ Young 1989, Kenney, van Gorkom, and Vollmer 2004). 
Thus, while the peak H$_2$ column density of NGC~4476 is a factor of a few
higher than typical values from Kenney \& Young's sample -- the
exact factor is resolution dependent, of course --
the HI column density in NGC~4476 is at least a factor of 
five lower than in the spirals.

Low HI column densities are also seen in the centers of early-type
spirals (Roberts \& Haynes 1994; van Driel \& van Woerden 1991).
Perhaps, then, HI is not detected in the center of NGC~4476 because
it has (or had, prior to stripping) a central HI hole like those in 
the early-type spirals.  It is worth noting, however, that HI maps of
low-luminosity ellipticals and E/S0s usually show centrally peaked HI
distributions, not central holes 
(Lake et al.\ 1987; Oosterloo, Morganti, \& Sadler 1999;
Sadler et al.\ 2000).
It would certainly be valuable to compare 
HI and H$_2$ maps of other gas-rich ellipticals to
see if they also show low HI column densities in regions where H$_2$
is present.  We could then investigate the balance of the atomic and
molecular phases in these  galaxies.  

If the interstellar medium of NGC~4476 has atomic and molecular
hydrogen in equilibrium, we can use the observed molecular column
densities to estimate expected HI column densities.
For these estimates we apply the models of
photodissociation regions presented by Allen~\etal~(2004).  The
observed peak CO intensity is 12.4 Jy beam$^{-1}$ km s$^{-1}$ or 24.3
K km s$^{-1}$ (Young 2002).  This CO intensity can be consistent with
our nondection of HI ($<$ 10$^{20}$ cm$^{-2}$) if and only if the
molecular clouds' average density is greater than 10$^3$ cm$^{-3}$
\emph{and} the UV field G$_{\circ}$ $\leq$ 1 (in units where the
``Habing Field'' over 4$\pi$ steradians is G$_{\circ}$=1.7).  CO
observations of other spiral galaxies typically trace gas of
average densities in the range $10^2$ to $10^3$ cm$^{-3}$. Thus,
either the HI column density in NGC 4476 is low because the molecular
clouds have higher density and lower interstellar UV fields than are
common for spirals, or the HI column density is low because of a
recent ram pressure stripping event, or both.  The prediction of higher
than average molecular gas densities in NGC 4476 could be tested with
observations of higher density molecular tracers.


\section{The Environmental impact on the Star Formation rate on NGC 4476}

Studies of Virgo Cluster spirals show that these galaxies are
deficient in HI and have reduced star formation rates in their outer disks,
while the star formation in the inner disk is normal or enhanced.  It
has been suggested that the HI deficiency is due to the systematic
removal of gas from the outer parts of the galaxies 
through ram pressure stripping (van Gorkom 2004;
Koopmann $\&$ Kenney 2002; Vollmer 2003).  If this is
indeed the case, and gas-rich ellipticals must suffer ram pressure
stripping just as gas-rich spirals do,
then the star formation in stripped ellipticals
should be reduced as is observed for stripped disk  galaxies. 

The standard tracers of star formation activity, such as H$\alpha$
emission, are often detected in elliptical galaxies and are usually
attributed to AGN activity, post-AGB stars, or cooling from hot gas rather
than to star formation.
But in the case of NGC~4476, which is known to contain appreciable amounts
of molecular gas, we propose to use far-IR (FIR) luminosity as an indicator
of star formation activity.
The star formation
rate (SFR) is calculated using the method derived by Condon, Cotton,
and Broderick (2002): 
\begin{equation}
\left(\frac{L_{FIR}}{L_\odot}\right)\sim~6.1\times10^{10} \left[\frac{SFR (M \geq 0.1 M_\odot)}{M_\odot~yr^{-1}}\right]
\end{equation}
The IRAS 60 $\micron$ and 100 $\micron$ fluxes of NGC 4476 (Knapp et al.\
1989) give
a FIR luminosity 4.3$\times$10$^8$ L$_\odot$ and a star formation rate
0.2~$M_\odot$~yr$^{-1}$ in stars with masses greater than
0.1~$M_\odot$.  If we divide the H$_2$ mass by the star formation
rate we obtain a depletion timescale of approximately 5$\times$10$^8$
years. This timescale is very similar to those of the
other CO-rich early type galaxies in Young's (2002) sample  
($\sim$ 3--10$\times$10$^8$ years;
Lucero $\&$ Young 2004 in preparation). 
The other
galaxies of that sample are field galaxies, so there is no evidence
here that the star formation efficiency in NGC~4476 has been affected by
ram pressure stripping.  

One may also attempt to measure a star formation rate from the radio
continuum emission of a galaxy, if the emission is not produced by an AGN
(Miller \& Owen 2002; Condon 1992).
But as mentioned in Section \ref{4476}, a deep VLA
image of M87 gives only an upper limit of 0.5 mJy for the flux density of
NGC~4476 at 20cm (Owen 2001, private communication).  
This limit and the IRAS fluxes give a logarithmic 
FIR/radio flux density ratio ($q$ value; Condon, Anderson, \& Helou
1991) of $q > 3.3$.  The average $q$ value for star forming galaxies is 2.3
with a dispersion of about 0.15 (Yun, Reddy, \& Condon 2001), so that the
FIR/radio flux density ratio in NGC 4476 is a factor of 10 higher than is
typical for star forming galaxies.  
It is well known that $q$ values tend
to increase at low luminosities (Yun et al.\ 2001; Condon et al.\ 1991) 
but the $q$ value for NGC~4476 is higher than any found in the Yun et al.\
(2001) sample.
Of course, the theoretical basis for the radio-FIR correlation (and for its
change in slope at low luminosities) has not been
firmly established.   



\section{Conclusions}
In this paper we present high dynamic range HI observations of the
low-luminosity elliptical galaxy NGC 4476.  In  summary:\\\\
$\bullet$ The spectral dynamic range of the final image (estimated as the
peak continuum intensity in the field divided by the rms noise level in
spectral channels) is $\sim$~50,000.\\  \\
$\bullet$ Our observations yield an upper limit to the HI mass of the
galaxy of approximately  $<$~10$^{7}$~$\mathrm{M_\Sun}$. \\ \\
$\bullet$ Assuming a minimum radius equal to that of the CO disk
radius, we obtain an HI surface density of  
 $\leq$ 1.5~$\mathrm{M_\Sun}$~pc$^{-2}$. \\ \\
$\bullet$ The ratio of the molecular gas mass to atomic gas mass in
the galaxy is $>$ 7, which puts NGC 4476 among the most seriously HI
deficient (compared to H$_2$) cluster  galaxies. \\\\ 
$\bullet$ An analysis of these new HI data suggest that the HI in NGC
4476 may have been removed due to ram-pressure stripping as the galaxy
traveled on a highly radial orbit through the Virgo Cluster core. It is
puzzling that there is no HI left in the inner disk, unless the
physical conditions in the molecular ISM of NGC 4476 are very different
(about a factor of 10 higher average density, coupled with
lower UV fields)  from
those of typical Virgo spirals.  

\acknowledgments
 We would like to thank Tim Cornwell for his input on the HI data
 reduction. This research has made use of the NASA/IPAC Extragalactic
 Database (NED) which is operated by the Jet Propulsion Laboratory,
 California Institute of Technology, under contract to the national
 Aeronautics and Space Administration. This work was partially
 supported by NSF grant AST-0074709 to New Mexico Institute of Mining
 and Technology and NSF grant AST-0098249 to Columbia  University. 
\clearpage
\bibliographystyle{apj3}
{\footnotesize
     \bibliography{lucero}
     } 
\clearpage
\begin{figure}
\plotone{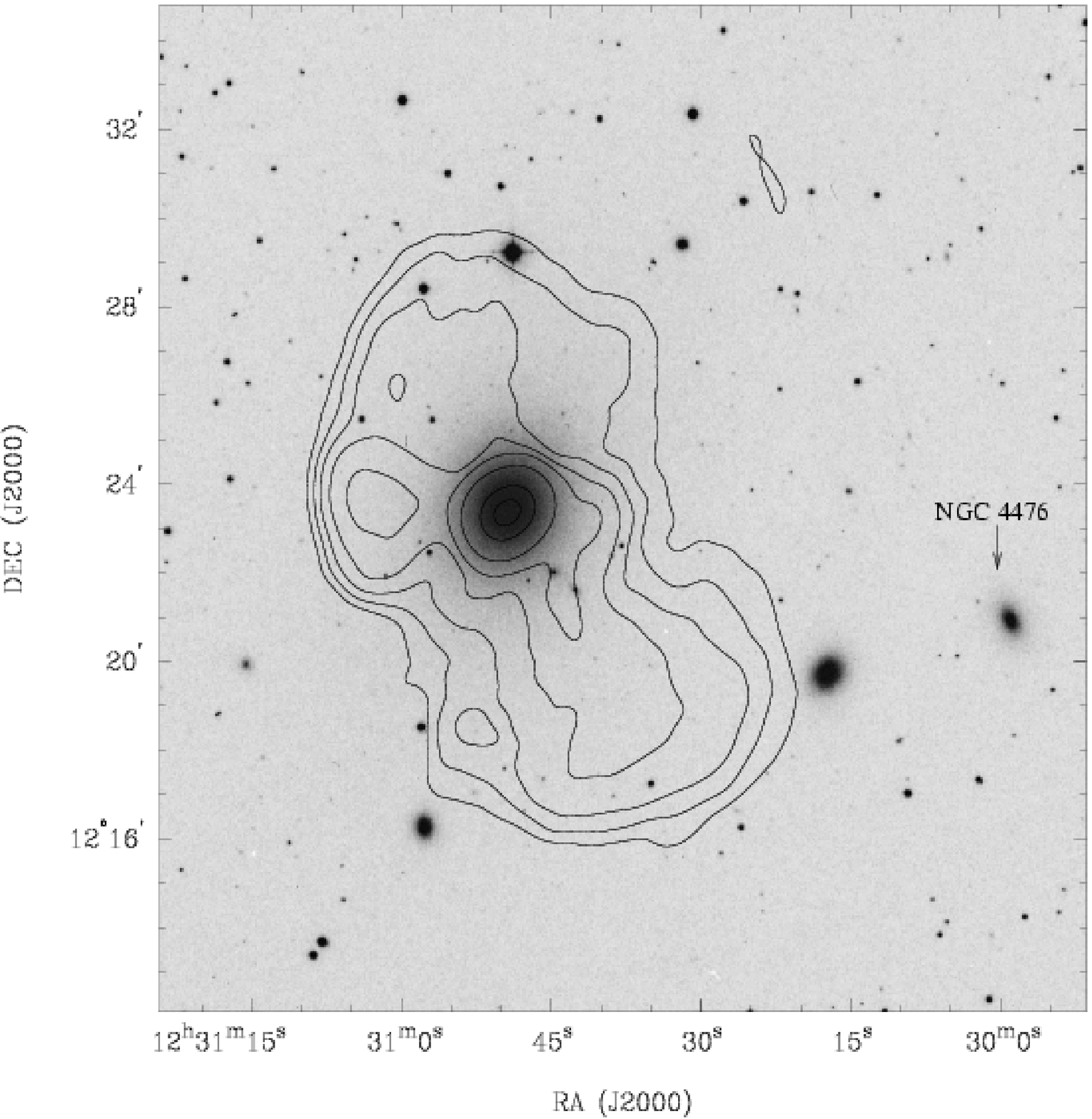}
\caption{Radio Continuum from channel zero data superposed on an optical
image from the digital sky survey.  The peak intensity is ~80
Jy~beam$^{-1}$, and the total flux density is 195 Jy.  Black contours show
the continuum intensity in units of 0.1$\%$, 0.3$\%$, 0.5$\%$, 0.8$\%$,
1.6$\%$, 5$\%$, 30$\%$, and 80$\%$ of the peak.  The dynamic range of this
image is $\sim$ 80,000.
\label{fig1}}
\end{figure} 
\clearpage
\begin{figure}
\plotone{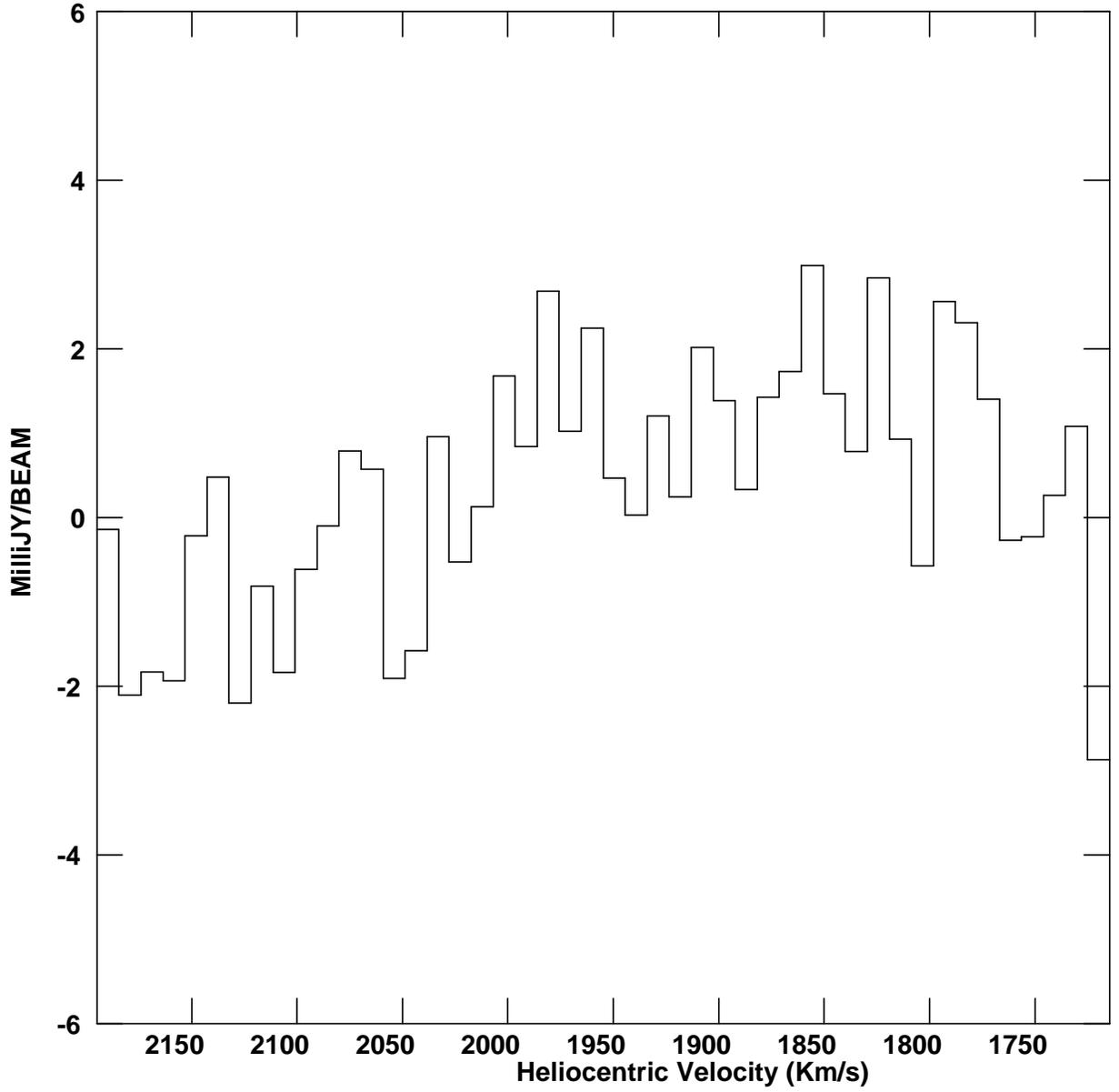}
\caption{Primary beam corrected HI spectrum extracted from one pixel across the entire usable velocity range (500 km s$^{-1}$) centered on NGC 4476.  The intensity (integrated over velocity) is 0.015$\pm$0.072 Jy km s$^{-1}$ in a 200 km s$^{-1}$ channel centered on 1960 km s$^{-1}$. \label{fig2}}
\end{figure} 
\clearpage
\begin{figure}
\plotone{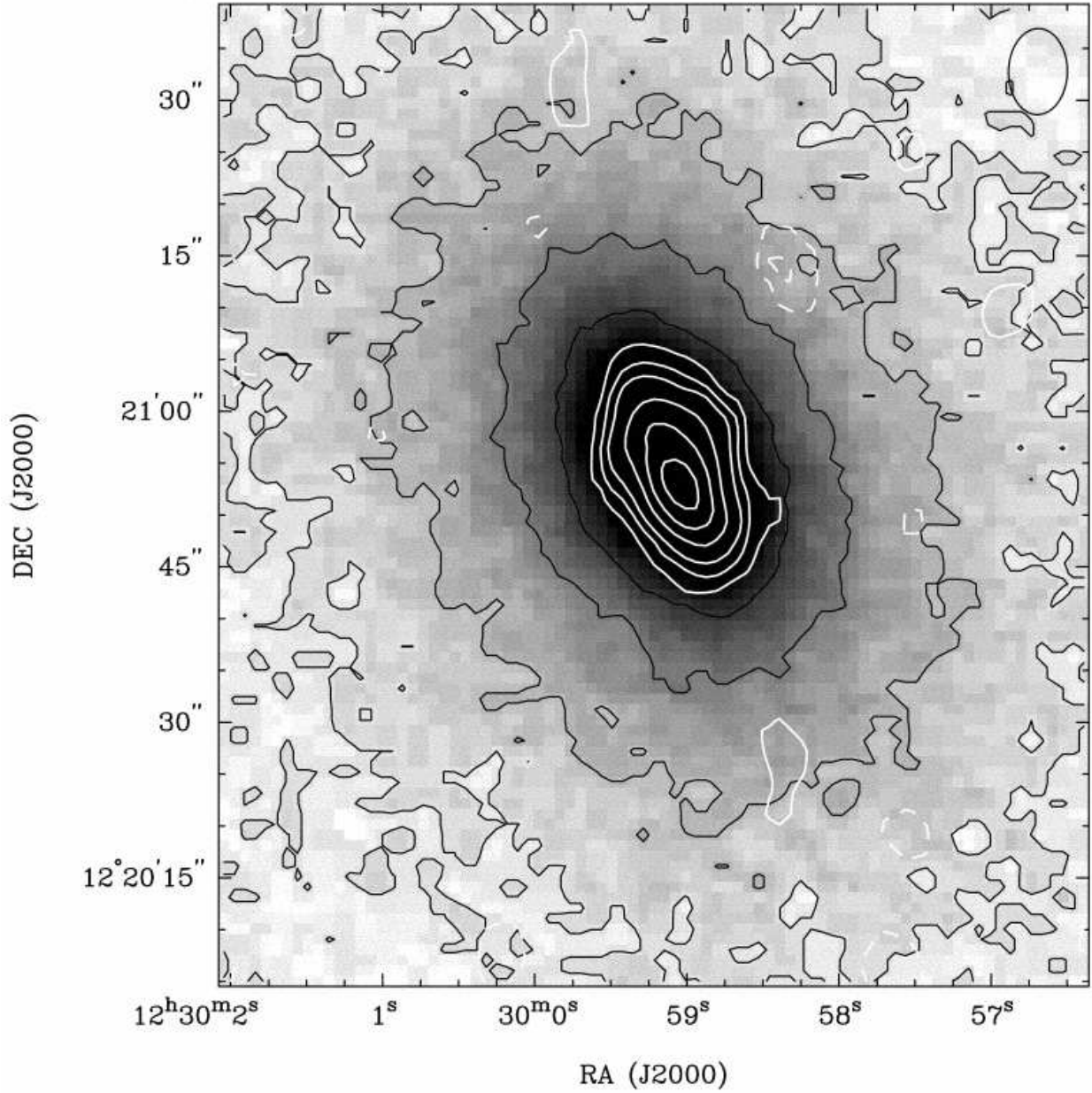}
\caption{CO emission (Young 2002). Heavy white contours show the CO integrated intensity in units of $-$20$\%$,$ -$10$\%$, 10$\%$, 20$\%$, 30$\%$, 40$\%$, 50$\%$, 70$\%$, and 90$\%$ of the peak (12.4 Jy beam$^{-1}$ km s$^{-1}$ $=$ 7.3$\times$10$^{21}$ cm$^{-2}$). \label{fig3}}
\end{figure}  
\clearpage
\begin{deluxetable}{lccccccl}
\tabletypesize{\scriptsize}
\tablecaption{Characteristics of Virgo Cluster Spirals with H$_2$/HI Ratios $\geq$ 10 . \label{tbl-1}}
\tablewidth{0pt}
\tablehead{
\colhead{Galaxy}   & \colhead{Type}    &\colhead{RA} &\colhead{DEC}  & \colhead{V$_{helio}$}     & \colhead{H$_2$/HI}  & \colhead{M$_{B}$} & \colhead{Environment}\\
\colhead{} & \colhead{} &\colhead{(J2000)} &\colhead{(J2000)}  & \colhead{km~s$^{-1}$}  & \colhead{} & \colhead{}  & \colhead{} 
}
\startdata
NGC 4388 &Sab    &12h25m46.7s &12d39m44s &2525   &10$\tablenotemark{a}$    &$-$19.37    &1.3 degrees from M87.\\
NGC 4647 &Sc     &12h43m32.3s &11d34m55s &1422   &20.6$\tablenotemark{a}$  &$-$19.19    &3.4 degrees from M87.\\
NGC 4419 &Sa     &12h26m56.4s &15d02m51s &$-$261 &25.3$\tablenotemark{b}$  &$-$18.57    &1.0 degrees from core of subcluster A.\\ 
NGC 4459 &SO/Sa  &12h29m00.0s &13d58m43s &1210   &100$\tablenotemark{c}$   &$-$19.72    &2.2$'$ away from companion, 1.6 degrees from M87.\\
NGC 4526 &SO/Sa  &12h34m02.2s &07d41m31s &448    &50$\tablenotemark{c}$    &$-$20.48    &5 degrees from M87\\
NGC 4710 &SO/Sa  &12h49m38.9s &15d09m56s &1125   &30$\tablenotemark{d}$    &$-$19.22    &6 degrees from M87
\enddata
\tablenotetext{a}{Knapp \etal~1987}
\tablenotetext{b}{Kenney $\&$ Young 1989}
\tablenotetext{c}{Thronson \etal~1989}
\tablenotetext{d}{Wrobel $\&$ Kenney 1992}
\tablecomments{Galaxy classifications are as given by individual authors.}
\end{deluxetable}

\begin{deluxetable}{lccccccl}
\tabletypesize{\scriptsize}
\tablecaption{Characteristics of Elliptical galaxies with H$_2$/HI Ratios $\geq$ 3. \label{tbl-2}}
\tablewidth{0pt}
\tablehead{
\colhead{Galaxy}   & \colhead{Type}    &\colhead{RA} &\colhead{DEC}  & \colhead{V$_{helio}$}     & \colhead{H$_2$/HI}  & \colhead{M$_{B}$} & \colhead{Environment}\\
\colhead{}  & \colhead{} &\colhead{(J2000)} &\colhead{(J2000)}   & \colhead{$km~s^{-1}$}  & \colhead{}  & \colhead{}  & \colhead{} 
}
\tablecolumns{8}
\startdata
NGC 2623 &E/Ir &08h38m24.080s &25d45m16.90s &5535 &3.0\tablenotemark{c} &$-$25.43 &member of a small group.\\
& & & & & & &Merger remnant, small radio core.\\
\cutinhead{HI nondetections}
NGC 4476 &E/SO  &12h29m59.1s & 12d20m55s   &1978    &$>$4.0\tablenotemark{a}     &$-$18.17     &12\arcmin from M87.\\
NGC 1275 &E     &03h19m48.1s &41d30m42s    &5264    &$>$3.4\tablenotemark{a}     &$-$21.50      &Perseus cluster member. \\
& & & & & & &Known to have AGN.\\

NGC 2783 &E7    &09h13m39.4s &29d59m35s    &6745    &$>$3.4\tablenotemark{b}     &$-$19.80     &Member of a small compact group\\
NGC 7052 &E4    &21h18m33.0s &26d26m49s    &4672    &$>$3.2\tablenotemark{b}     &$-$20.30     &Field galaxy with no nearby companion.\\           &            &             &        &           &                 &     &Known to have AGN.\\
\cutinhead{H$_2$ nondetections}
Haro 20 &E/Ir &03h28m14.532s &$-$17d25m10.49s &1866 &$<$57.6\tablenotemark{a} &$-$16.3 &Field galaxy.\\
NGC 3226 &E2 &10h23m27.000s &19d53m54.36s &1151 &$<$56\tablenotemark{a} &$-$18.2 &Member of Interacting pair, AGN.\\
NGC 3773 &E1/SO &11h38m12.97s &12d06m42.9s &987 &$<$6.7\tablenotemark{a} &$-$17.0 &Merger remnant\\
NGC 5018 &E3 &13h13m00.99s &$-$19d31m05.1s &2794 &$<$3.8\tablenotemark{a} &$-$19.3 &Field galaxy, non-interacting\\
 &            &             &        &           &                 &     &companion.\\
NGC 5363 &E3p/IO &13h56m07.24s &05d15m17.0s &1139 &$<$9.5\tablenotemark{a} &$-$19.2 &Field galaxy with non-interacting\\ 
 & & & & & & &companion\\
IC 1182 &EOp/SO &16h05m36.798s &17d48m07.70s &10240 &$<$43\tablenotemark{a} &$-$20 &Known to have AGN\\
\enddata
\tablenotetext{a}{Lees, Knapp, Rupen, and Phillips 1991}
\tablenotetext{b}{Wiklind, Combes, and Henkel 1995}
\tablenotetext{c}{Georgakakis \etal~2002}
\tablecomments{Galaxy classifications are as given by individual authors.}
\end{deluxetable}
\clearpage
\begin{table}
\caption{Properties of NGC 4476\label{tbl-3}}
\begin{tabular}{lr}
\tableline\tableline 
Other names                   &UGC 07637 \\
                              &VCC 1250\\
                              &CGCG  070-128\\
$\alpha$ (J2000)              &12$^h$29$^m$59.1$^s$\\
$\delta$ (J2000)              &12$^d$20$^m$55$^s$\\
Morphological type\tablenotemark{a}            &E/SO\\
Distance to M87 (arcmin)      &12\\
Heliocentric velocity (km/s)\tablenotemark{b}  &1978$\pm$~12\\
Distance D (Mpc)              &17$\pm$1.4\\
H$_2$ mass (M$_\odot$)\tablenotemark{c}    & $1.1\times 10^8$\\
L$_{FIR}$, L$_\odot$  &  4.3$\times 10^8$ \\
L$_{1.4 GHz}$, W m$^{-2}$  & $< 2\times 10^{19}$ \\
\tableline
\end{tabular}
\tablenotetext{a}{Knapp \etal 1989}
\tablenotetext{b}{Simien $\&$ Prugniel 2002}
\tablenotetext{c}{Young 2002}
\tablecomments{Coordinates are for epoch J2000.}
\end{table}
\clearpage
\begin{table}
\caption{Gas properties of the ISM and IGM.\label{tbl-4}}
\begin{tabular}{lc}
\tableline\tableline 
$R_{CO}~(kpc) $                                &1.2\\
\tablenotemark{a} $n_{e}$~(cm$^{-3}$)     &3.6$\times$10$^{-3}$\\
\tablenotemark{b} $V_{rot}$~(km~s$^{-1}$)      &105\\
$V_{gal}$~(km~s$^{-1}$)                        &671\\ 
$P_{ram}$~(N~m$^{-2}$)                         &3.1$\times$10$^{-12}$\\
$\Sigma_{gas}^{H_2}$~($M_{\odot}$~pc$^{-2}$)   &24.3\\
N$_{HI}$ cm$^{-2}$                             &\\
...3$\sigma$ integrated limit:                 &$<$ 1.2$\times$10$^{20}$\\
...6$\sigma$ individual channel limit:         &$<$ 5.4$\times$10$^{19}$\\
$\Sigma_{gas}^{HI}$~($M_{\odot}$~pc$^{-2}$):   &\\
...3 $\sigma$ integrated limit:                &$<$1.0\\
...6 $\sigma$ individual channel limit:        &$<$0.4\\
M(H$_2$)/M(HI):                                      &\\
...3 $\sigma$ integrated limit:                &$>$~7\\
...6 $\sigma$ individual channel limit:        &$>$~16\\
Restoring Pressure for HI:                     &\\
P$^{HI}_{rest}$~(N~m$^{-2}$):                  &\\
...3 $\sigma$ integrated limit:                &$<$~6.0$\times$10$^{-13}$\\
...6 $\sigma$ individual channel limit:        &$<$~2.7$\times$10$^{-13}$\\
Restoring Pressure for H$_2$:                  &\\
P$^{H_2}_{rest}$~(N~m$^{-2}$)                  &1.5$\times$10$^{-11}$\\
Star Formation Rate $\mathrm{M_\Sun}~yr^{-1}$  &0.2\\
\tableline
\end{tabular}
\tablenotetext{a}{Density of the ICM at the projected distance of 60 kpc,
given by Nulsen $\&$ B\"ohringer 1995.}
\tablenotetext{b}{Rotational velocity of the CO disk (Young 2002). We adopt this velocity as the HI rotational velocity.}
\end{table}
\end{document}